\documentclass[11pt]{article}
\usepackage{hyperref}
\usepackage{amsmath}
\usepackage{amssymb}
\newcommand{\la}{\label}
\newcommand{\bbm}{\begin{multline}}
\newcommand{\eem}{\end{multline}}
\newcommand{\be}{\begin{equation}}
\newcommand{\ee}{\end{equation}}
\newcommand{\bea}{\begin{eqnarray}}
\newcommand{\eea}{\end{eqnarray}}
\newcommand{\p}{\partial}

\newcommand{\comment}[1]{}
\usepackage{color}

\renewcommand{\title}[1]{%
    \bigskip%
    \begin{center}%
    \Large\bf #1%
    \end{center}%
    \vskip .2in}

\renewcommand{\author}[1]{%
    {\begin{center}
    #1
    \end{center}}}
\newcommand{\address}[1]{\vspace{-1.7em}\vspace{0pt}
    {\begin{center}
    \it #1
    \end{center}}}

\begin{document}


\title{Comment on ``Thermal Hall Effect and Geometry with Torsion"}

\author
{
Rabin Banerjee  $\,^{\rm a,b}$,
Pradip Mukherjee $\,^{\rm c,d}$}
\address{$^{\rm a}$S. N. Bose National Centre 
for Basic Sciences, JD Block, Sector III, Salt Lake City, Kolkata -700 098, India }

\address{$^{\rm c}$Department of Physics, Barasat Government College,\\Barasat, West Bengal

 }

\address{$^{\rm b}$\tt rabin@bose.res.in}
\address{$^{\rm d}$\tt mukhpradip@gmail.com}

\begin{abstract}
This is a comment on a paper by Gromov and Abanov [Phys. Rev. Lett. 114, 016802 (2015)]. We will show that there is an inconsistency which renders the results untenable.
\end{abstract}

This is a comment on a paper by Gromov and Abanov \cite{GA}. We will show that there is an inconsistency which renders the results untenable. In their paper \cite{GA} a complex scalar field theory which is invariant under space and time translations is coupled with background curvature in the torsional Newton - Cartan (NC)formulation. They introduced the vielbeins ${E_a}^\mu$ where $a =0, A; A=1,2$ and $\mu = 0, i; i=1,2$, where $a$ and $\mu$ respectively label the tangent space and the curved space coordinates. The derivatives with respect to these coordinates are related by appropriate vielbeins 
${E_a}^\mu$
\be
 \la{vielbein}
	\p_a \rightarrow {E_a}^\mu {\partial_\mu},
\ee
The basis one-forms ${e_\mu}^a $ are then defined
so that
\begin{eqnarray}
\la{oneforms}
{e_\mu}^a{E_b}^\mu = \delta^a_b, \quad {e_\mu}^a{E_a}^\nu = \delta^\mu_\nu
\end{eqnarray}
The NC geometry is charcterized by the authors by constructing 
a degenerate metric 
\be 
\la{metric}h^{\mu\nu} = \delta^{AB}{E_A}^\mu{E_B}^\nu
\ee
and an  1-form by
\be
\la{oneform1}
n_\mu = e^0_\mu.
\ee
They are defined to satisfy the NC geometry requirements
\be
 \la{NCprop}
	{E_0}^\mu n_\mu = 1, \quad h^{\mu\nu}n_\nu = 0.
\ee


After defining the NC geometry in the general way by equations (\ref{metric}), (\ref{oneforms}) and (\ref{NCprop}) from the vielbines the authors go on to assume that in a particular parametrization
 the metric takes the following form:
\be
h^{\mu\nu}=\left(\begin{array}{cc}
\frac{n^2}{n_0^2} & -\frac{n^i}{n_0} \\
 -\frac{n^i}{n_0} & h^{ij}\\
\end{array}\right),
\la{h}
\ee
where $n^i=h^{ij}n_j$ and $n^2 = n_i n_j h^{ij}$.  It may be observed that the authors are considering NC spacetime to be foliated in spacelike hypersurfaces where $h^{ij}$ is a nondegenerate
metric atributed to the hypersurface. In other words,the authors are using the Galilean frame. This is supported by their declaration -- ``Notice, that the
spatial part of the metric $h^{ij}$ is a (inverse) metric on a
fixed time slice, it is symmetric and invertible."

We first observe that for the NC spacetime subject to metric compatibility,
\begin{equation}\label{new1}
n_\mu = \p_\mu t,
\end{equation}
for some time function $t$. This has nothing to do with torsion or no torsion. To see this note that due to metric compatibility, $\nabla_\rho\tau_\mu = 0$, where
\begin{equation}
\nabla_\rho\tau_\mu = \partial_\rho\tau_\mu +\Gamma^{\lambda}_{\rho\mu}\tau_
\lambda
\end{equation}
is the covariant derivative and $\Gamma^{\lambda}_{\rho\mu}$ is the connection. On account of the metric compatibility,
\begin{equation}
\nabla_\rho\tau_\mu - \nabla_\mu\tau_\rho = 0\label{P}
\end{equation}
Then, on account of the Poincare lemma there exists (at least locally)
a function $t$ of the spacetime coordinates such that,
\begin{equation}
n_\mu = \nabla_\mu t=\p_\mu t
\end{equation}
In case of the torsional NC geometry. (\ref{new1}) is only locally true. But that is sufficint for us.

The NC spacetime can be foliated in spacelike hypersurfaces in a unique way, using t as the affine parameter \cite{K,J}. In case of standard (torsionless) NC geometry t is a global function. Otherwise it is local. The ``fixed time slice" considered by the authprs is every where orthogonal to $n_\nu$. On it is defined the nondegenerate spatial metric $h^{ij}$.The contravariant and covariant components of $3$ vectors can be related as
\begin{eqnarray}\label{V}
^{(3)}V^i = h^{ij}V_j\la{3} 
\end{eqnarray}
In other words we have to choose 
\be x^0 =t .
\la{adapted}
\ee 
as our coordinate time. This means working in the adapted coordinates \cite{abpr} \cite{K,J}. From \eqref{oneform1} and \eqref{new1} we obtain
\begin{equation}\label{new2}
e_\mu^0 = \partial_\mu t = e_\mu^b \partial_b t,
\end{equation}
which leads to \eqref{adapted}. Recalling that the  spacelike hypersurface is orthogonal to the direction of local time flow, we obtain, from the above relations, the following parametrisation,
\begin{equation}\label{new4}
{e_\mu}^0 = (1,0,0,0).
\end{equation}

Our point is now to show that the above results are inconsistent with the choice of metric \eqref{h}. A simple calculation is sufficient to prove it. From the above definitions we find
\begin{equation}\begin{aligned}
n^2 &= n_in_jh^{ij} = n_in_jE_A^{{}i} E_A^{{}j} = (n_iE_A^{{}i})(n_jE_A^{{}j})\\
                   &= (n_{\mu}E_A^{{}\mu} - n_0E_A^{{}0})(n_{\nu}E_A^{{}\nu} - n_0E_A^{{}0})\\
                   &= n_\mu n_\nu h^{\mu\nu} - 2n_0 E_A^{{}0} n_{\mu} E_A^{{}\mu} + (n_0 E_A^{{}0})^2 \\
                   &=  - 2n_0E_A^{{}0} n_{\mu}E_A^{{}\mu} + (n_0E_A^{{}0})^2
                   \la{nsquare}\end{aligned}
\end{equation}
where we have used (\ref{metric}) and (\ref{NCprop}).

Now, on the other hand, (\ref{h}) and (\ref{metric}) give

\be 
n^2 = n_0^2 h^{00} = (n_0E_A^{{}0})^2.
\la{next}
\ee

Combining (\ref{nsquare})and (\ref{next}) we are left with two possibilities:
\begin{enumerate}
\item 
$n_{\mu}E_A^{{}\mu} =0 $, or
\item $ n_0E_A^{{}0} = 0$
\end{enumerate}

The first condition is superfluous since it is identical to the second one. To see this we use \eqref{oneform1} and \eqref{new4} to find, 
\begin{equation}\label{new3}
n_{\mu}E_A^{{}\mu} = {e_\mu}^0E_A^{{}\mu}=E_A^{{}0} = 0.
\end{equation}
On the other hand the second condition also leads to the same result, since $n_0$ is non-vanishing.

As a consequncee $h^{00} = 0$
and $n^2 = 0$. This means in turn $n^i =0$. Thus $h^{0i} =h^{i0} = 0$. These conclusions follow in order to satisfy the second condition in (\ref{NCprop}). We are thus led to the following structure of the metric
 \begin{equation}\label{eq10}
h^{\mu\nu}=\left(\begin{array}{cc}
0 & 0 \\
 0 & h^{ij}\\
\end{array}\right).
\end{equation}
This metric has appeared in earlier studies \cite{bmm, abpr} on NC geometry.

In fact, that the construction of the metric in \cite{GA} leads to (\ref{eq10}) can be understood from much more simple considerations. Look at the vector $n^i$ on the 'fixed time slice'. We construct
$\xi^\mu = (0, n^i)$. It is evident that $\xi^\mu$ is a space like vector. Then it is also undeniable that $\xi^\mu = h^{\mu\nu}\lambda_\mu$, where, $\lambda_\mu = (0, n_i)$. From the metric properies of the NC geometry,
$h^{\mu\nu}\lambda_\mu\lambda_\nu =0$ i.e $n^2 = h^{ij}n_in_j = 0$. The $(0,0)$ element of the authors proposed matrix vanishes. Again, since $h^{ij}$ is invertible, $n^2=0$ implies $n_i=0$.

To conclude, we show that the structure (\ref{h}) is inconsistent with the particular parametrization chosen in \cite{GA}. Consistency is recovered by the choice \ref{eq10}. The key element is that the various structures assumed by the authors lead to the adapted coordinate $x^0=t$, where $t$ defines a Galilean frame \cite{K,J}.

Since the subsequent analysis in \cite{GA} is done by using the metric \eqref{h}, the results become untenable. Naturally, the physical conclusions drawn from the calculations based on (\ref{h}) are questionable. .

\end{document}